\documentclass[12pt]{article}
\usepackage{amsmath,latexsym, amssymb}
\usepackage{epic}
\usepackage{graphics}
\usepackage{graphicx}

\numberwithin{equation}{section}

\usepackage{color}
\usepackage{epigraph}
\begin{document}

%
%
%
%
%
%
%
%
%

\title{Exact and "exact" formulae in the theory of composites}

\author{Igor Andrianov$^1$, Vladimir Mityushev$^2$\\
$^1$RWTH Aachen University, Germany  \\
igor\_andrianov@hotmail.com\\
$^2$Pedagogical University\\ 
Krakow 30-084, Poland mityu@up.krakow.pl}




\date{}
\maketitle
\begin{abstract}
The effective properties of composites and review literature on the methods of Rayleigh, Natanzon--Filshtinsky, functional equations  and asymptotic approaches are outlined.  In connection with the above methods and new recent publications devoted to composites, we discuss the terms {\it analytical formula, approximate solution, closed form solution, asymptotic formula} etc... frequently used in applied mathematics and engineering in various contexts. Though mathematicians give rigorous definitions of exact form solution the term "exact solution" continues to be used too loosely and its attributes are lost. In the present paper, we give examples of misleading usage of such a term. 
\end{abstract}\newpage

--------------------------------------------------------------------
\epigraph{{\it Unfairly recognizes advertising, which advertises under the guise of one commodity another product.}}{--- a shortened form of FTC Act, as amended, \S 15(a), 15 U.S.C.A. \S 55(a) (Supp. 1938)} 

\epigraph{{\it The complexity of the model is a measure of
misunderstanding the essence of the problem.}}{--- A.Ya. Findlin \cite{Findlin}} 

\section{Introduction}
\label{sec:Int} 
This paper is devoted to the effective properties of composites and review literature concerning analytical formulae. 
The terms {\it analytical formula, approximate solution, closed form solution, asymptotic formula} etc... are frequently used in applied mathematics and engineering in various contexts. Though mathematicians give rigorous definitions of exact form solution \cite{Gakh77, MiR, GMN} 
the term "exact solution" continues to be used too loosely and its attributes are lost. In the present paper, we give examples of misleading usage of such a term. This leads to paradoxical situations when an author $A$ approximately solves a problem and an author $B$ repeats the result of $A$ adding the term "exact solution". Further, the result of $B$ is dominated in references due to such an exactness. Examples of such a wrongful usage of the exactness terms are given in the present paper.          

We use also the term {\it constructive solution (method)} understood from the pure utilization point of view. The constructive solution in the present paper means that we have a formula or a precisely described algorithm, perhaps, on the level of symbolic computations. Problems are said to be {\it tractable} if they can be solved in terms of a closed form expression or of a constructive method. For instance, a typical result from the homogenization theory consists in replacement of a PDE with high oscillated coefficients by an equation with constant coefficients called the effective constants. Next, a boundary value problem is stated to determine the effective constants. Therefore, the homogenization theory rigorously justifies existence of the effective properties. Their determination is a separate question concerning a boundary value problem in a periodicity cell. Though, in some cases, we should not solve a boundary value problem (see for instance the Keller-Dykhne formula \cite{Dykhne}).

We think that usage of the terms "constructive solution" and more strong "exact formula" are not acceptable when one writes a formula when its entries could be found from an additional numerical procedure. One may use its own terms 
but he/she has to be consistent with the commonly used terminology in order to avoid misleading. We want to exclude situations when researchers should necessary add that "our formula is really exact" since it is a spiral way and in the next time they should add that "our formula is really--really exact". Below, we try to discuss the levels of exactness and precise the clear for engineers terminology which should be used in applied mathematics.

The solution of an equation $A x=b$ can be formally written in the form $x=A^{-1}b$ where $A^{-1}$ is an inverse operator to $A$. 
 
We say that $ x=A^{-1}b$ is a {\it analytical form solution} if the expression $A^{-1}b$ consists of a finite number of elementary and special functions, arithmetic operations, compositions, integrals, derivatives and  series.  {\it Closed form solution} usually excludes usage of series \cite{Gakh77,MiR,GMN}\footnote{However, an integral is accepted. In the same time, Riemann's integral is a limit of Riemann's sums, i.e., it is a series.}.

{\it Asymptotic methods} \cite{Andrianov1,Andrianov2}  are assigned to analytical methods when solutions are investigated near the critical values of the geometrical and physical parameters. Hence, asymptotic formulae can be considered as analytical approximations.

In order to distinguish our results from others we proceed to classify different types of solutions.

{\it Numerical solution} means here the expression $x=A^{-1}b$ which can be treated only numerically. {\it Integral equation methods} based on the potentials of single and double layers usually give such a solution. An integral in such a method has to be approximated by a cubature formula. This makes it pure numerical, since cubature formulae require numerical data in kernels and fixed domains of integrations. 

{\it Series method} arises when an unknown element $x$ is expanded into a series $x = \sum_{k=1}^{\infty}c_k x_k$ on the basis $\{x_k\}_{k=1}^{\infty}$ with undetermined constants $c_k$.  Substitution of the series into equation can lead to an infinite system of equations on $c_k$. In order to get a numerical solution, this system is truncated and a finite system of equations arise, say of order $n$. Let the solution of the finite system tend to a solution of the infinite system, as $n\to \infty$. Then, the infinite system is called regular  and can be solved by the described {\it truncation method}. This method was justified for some classes of equations in the fundamental book \cite{KK}. The series method can be applied to general equation $A x=b$ in a discrete space in the form of infinite system with infinite number of unknowns. In particular, Fredholm's alternative and the Hilbert-Schmidt theory of compact operators can be applied \cite{KK}. So in general, the series method  belongs to numerical methods. In the field of  composite materials, the series truncation method was systematically used by Guz et al. \cite{Guz} and many others.


The special structure of the composite systems or application of a low-order truncation can lead to an approximate solution in symbolic form.  Par excellence examples of such solutions are due to Rayleigh \cite{Rayleigh} and to McPhedran et al. \cite{Mc, McPhedran1988, McPhedran1978} where analytical approximate formulae \label{page-RM} for the effective properties of composites were deduced.

{\it Discrete numerical solution} refers to applications of the finite elements and difference methods. 
These methods are powerful and their application is reasonable when the geometry and the physical parameters are fixed. Many experts perceive a pristine computational block (package) as an exact formula:  just substitute data and get the result!  However, a sackful of numbers is not as useful as an analytical formulae. Pure numerical procedures can fail as a rule for the critical parameters and analytical matching with asymptotic solutions can be useful even for the numerical computations. Moreover, numerical packages sometimes are presented as a remedy from all deceases. It is worth noting again that numerical solutions are useful if we are interested in a fixed geometry and fixed set of parameters for engineering purposes. 

Analytical formulae are useful to specialists developing codes for composites, especially for optimal design. We are talking about the creation of highly specialized codes, which enable to solve a narrow class of problems with exceptionally high speed. "It should be emphasized that in problem of design optimization the requirements of accuracy are not very high. A key role plays the ability of the model to predict how the system reacts on the change of the design parameters. This combination of requirements opens a road to renaissance of approximate analytical and semi-analytical models, which in the recent decades have been practically replaced by "universal" codes" \cite{Findlin}.

Asymptotic approaches allow us to define really important parameters of the system. The important parameters in a boundary value problem are those that, when slightly perturbed, yield large variations in the solutions. In other words, asymptotic methods make it possible to evaluate the sensitivity of the system. There is no need to recall that in real problems the parameters of composites are known with a certain (often not very high) degree of accuracy. This causes the popularity of various kinds of assessments in engineering practice. In addition, the fuzzy object oriented (robust in some sense) model \cite{Ferguson} is useful to the engineer. Multiparameter models rarely have this quality. "You can make your model more complex and more faithful to reality, or you can make it simpler and easier to handle. Only the most naive scientist believes that the perfect model is the one that represents reality" \cite[p. 278]{Gleik}. It should also be remembered that for the construction of multiparameter models it is necessary to have very detailed information on the state of the system. Obtaining such information for an engineer is often very difficult for a number of objective reasons, or it requires a lot of time and money. And the most natural way of constructing sufficiently accurate low-parametric models is to use the asymptotic methods \cite{Tayler}.


\section{Approximate and exact constructive formulae in the theory of composites}
\label{sec:met}
In the present section, we discuss constructive methods used in the theory of dispersed composites. The main results are obtained for 2D problems which will be considered in details.  

Let $\omega_1$ and $\omega_2$ be the fundamental pair of periods on the complex plane $\mathbb C$ such that $\omega_1>0$ and $\textrm{Im}\; \omega_2 > 0$ where $\textrm{Im}$ stands for the imaginary part. The fundamental parallelogram $Q$ is defined by the vertices $\pm \frac{\omega_1}2$ and $\pm \frac{\omega_2}2$. Without loss of generality the area of $Q$ can be normalized to one. 
The points $m_1 \omega_1+ m_2\omega_2$ ($m_1,m_2 \in \mathbb Z$) generate a doubly periodic lattice $\mathcal Q$ where $\mathbb Z$ stands for the set of integer numbers. Introduce the zero-th cell 
$$Q= Q_{(0,0)}=\left\{z=t_1 \omega_1+i t_2 \omega_2 \in  \mathbb{C}:-\frac{1}2<t_1,t_2<\frac{1}2\right\}.$$
The lattice $\mathcal{Q}$ consists of the cells
$Q_{(m_1,m_2)}=Q_{(0,0)}+m_1\omega_1+i m_2\omega_2$. 

\begin{figure}[!ht]
\centering 
\includegraphics[clip,
width=0.9\textwidth]{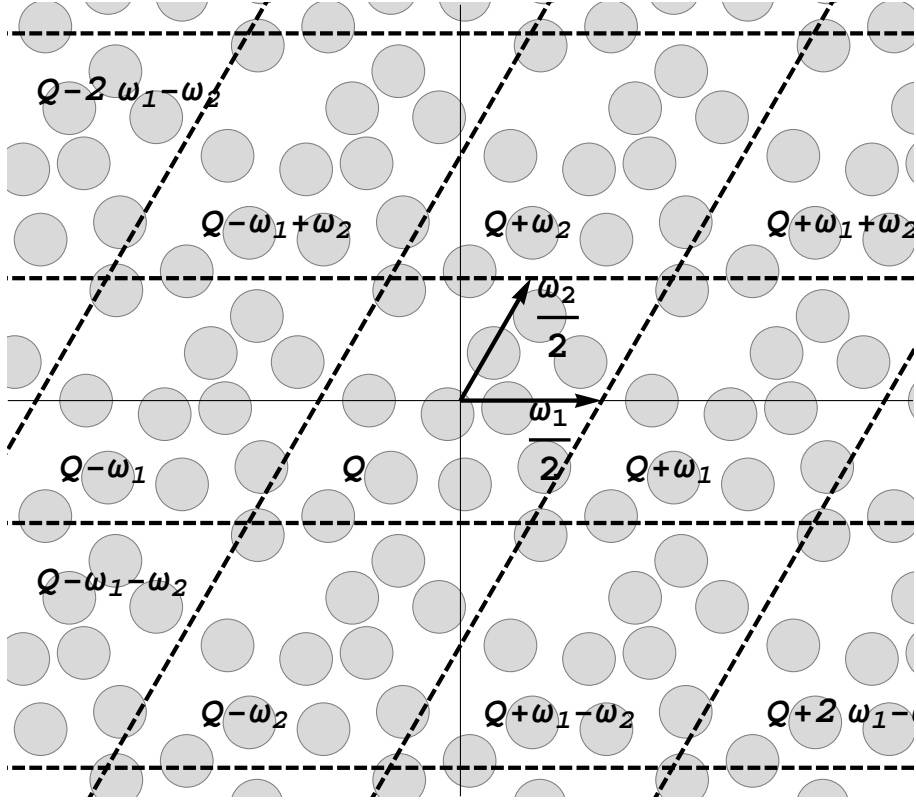}
\caption{Doubly periodic composite. 
}
\label{fig:DisksRandomPer1}
\end{figure}

Consider $N$ non--overlapping simply connected domains $D_k$ in the cell $Q$ with Lyapunov's boundaries $L_k$ 
and the multiply connected domain $D=Q \backslash \cup_{k=1}^N(D_k \cup L_k)$, the complement of all the closures of $D_k$ to $Q$  (see Fig.\ref{fig:DisksRandomPer1}). The case of equal disks $D_k$ will be discussed below.

We study conductivity of the doubly periodic composite when the host $D+m_1 \omega_1+ m_2\omega_2$ and the inclusions $D_k+m_1 \omega_1+ m_2\omega_2$ are occupied by conducting materials. Introduce the local conductivity as the function
\begin{equation}
\sigma(\mathbf x) = \left\{
\begin{array}{llll}
1,\quad \mathbf x \in D,
\\
\sigma, \quad \mathbf x \in D_k, \; k=1,2,\ldots,N.
\end{array}
\right.
\label{local}
\end{equation}  
Here, $\mathbf x=(x_1,x_2)$ is related to the above introduced complex variable $z$ by formula $z=x_1+ix_2$. The potentials $u(\mathbf x)$ and $u_k(\mathbf x)$ are harmonic in $D$ and $D_k$ ($k=1,2,\ldots,N$) and continuously differentiable in closures of the considered domains. The conjugation conditions express the perfect contact on the interface
\begin{equation}
u=u_k,\quad \frac{\partial u}{\partial \mathbf n}=\sigma\frac{\partial
u_{k}}{\partial \mathbf n}\quad\text{on } L_k=\partial D_k, \quad k=1,2,\ldots,N,
\label{eq:2.1}
\end{equation}
where $\frac{\partial}{\partial \mathbf n}$ denotes the outward normal derivative to $L_k$. The external field is modelled by the quasi--periodicity conditions 
\begin{equation}
u(z+\omega_1)=u(z)+\xi_1,\;u(z+\omega_2)=u(z)+\xi_2,  
\label{eq:quasi}
\end{equation}
where $\xi_{1,2}$ are constants modeled the external field applied to the considered doubly periodic composites. 
%

Consider the regular square lattice with one disk $|z|<r_0$ per cell. In this case,  $\omega_1=1$, $\omega_2 =i$ and $N=1$. The effective conductivity tensor is reduced to the scalar effective conductivity $\sigma_e$. In order to determine $\sigma_e$ it is sufficient to solve the problem \eqref{eq:2.1}-\eqref{eq:quasi} for $\xi_1=1$, $\xi_2=0$  and calculate 
\begin{equation}
\sigma_e = 1+ (\sigma-1)\;\pi r_0^2 \;\frac{\partial u_1} {\partial x_1} (0).
\label{e11b}
\end{equation}

\subsection{Method of Rayleigh}
The first constructive solution to the conductivity problem \eqref{eq:2.1}-\eqref{eq:quasi} for the regular square array was obtained in 1892 by Lord Rayleigh \cite{Rayleigh}. The problem is reduced to an infinite system of linear algebraic equations by the series method outlined below. 
The potentials are expanded into the odd trigonometric series in the local polar coordinates $(r, \theta)$
\begin{equation}
u_1(r, \theta) = C_0+C_1 r \cos \theta + C_3 r^3 \cos 3\theta+\ldots ,
\;r \leq r_0,
 \label{eq:Ray1}
\end{equation}
\begin{equation}
u(r, \theta) = A_0+(A_1 r+ B_1 r^{-1})\cos \theta+ (A_3 r^3+ B_3 r^{-3}) \cos 3\theta+\ldots.
 \label{eq:Ray2}
\end{equation}
The same series can be presented as the real part of analytic functions expanded in the Taylor and Laurent series, respectively,
\begin{equation}
\varphi_1(z) = \varphi_{10}+\varphi_{11}z + \varphi_{13}z^3+\ldots ,
\;r \leq r_0,
 \label{eq:Ray1a}
\end{equation}
\begin{equation}
\varphi(z)  = \alpha_0+\alpha_1 z+ \beta_1 z^{-1}+ \alpha_3 z^3+ \beta_3 z^{-3}+\ldots,
 \label{eq:Ray2a}
\end{equation}
where $u_1(z) =$Re $\varphi_1(z)$ and $u(z) =$Re $\varphi(z)$.
Substitution of the series \eqref{eq:Ray1}-\eqref{eq:Ray2} (or the series \eqref{eq:Ray1a}-\eqref{eq:Ray2a}) into \eqref{eq:2.1} and selection the terms on $\cos k\theta$ yields an infinite system of linear algebraic equations.  Rayleigh treats the quasi-periodicity condition \eqref{eq:quasi} as the balance of the multiple sources inside and out of the disk $D_1$. This leads to
the lattice sums $S_m=\sum_{m_1,m_2} (m_1+~im_2)^{-m}$ ($m=2,3,\ldots$), where $m_1$, $m_2$ run over all integers except $m_1 =m_2=0$. 
The series $S_2$ is conditionally convergent, hence, its value depends on the order of summation. Rayleigh uses the Eisenstein summation \cite{weil} 
\begin{equation}
\sideset{}{^e}\sum_{m_1,m_2} :=  \lim_{M_2\rightarrow +\infty} \lim_{M_1\rightarrow +\infty}
\sum_{m_2=-M_2}^{M_2} \sum_{m_1=-M_1}^{M_1}.  
\label{eq:s6}
\end{equation}
It is worth noting that Rayleigh (1892) did not refer to Eisenstein (1847) and used Weierstrass's theory of elliptic functions (1856). Perhaps it happened because Eisenstein treated his series formally without study on the uniform convergence introduced by Weierstrass later. Rayleigh used the Eisenstein summation and proved the fascinating formula $S_2=\pi$ for the square array (see discussion in \cite{Yakubovich}).   

The coefficients of the infinite system are expressed in terms of the lattice sums $S_m$. This system is written in the next section in the complex form \eqref{eq:ex9R4}. Rayleigh truncates the infinite system to get an approximate formula for the effective conductivity \eqref{e11b}. 

Rayleigh extended his method to  rectangular arrays of cylinders and to 3D cubic arrays of spherical inclusions. 
Rayleigh's approach was elaborated by McPhedran with coworkers \cite{Mc, McPhedran1988, McPhedran1978}. For instance, for the hexagonal array Perrins et al. \cite{Mc} obtained the approximate analytical formula
\begin{equation}
\label{proxr}
\sigma_e \approx 
1+ \frac{2 f \varrho}{1-f\varrho-\frac{0.075422 \varrho^2 f^6}{1-1.060283\varrho^2 f^{12}}-0.000076 \varrho^2 f^{12}},
\end{equation}
where $\varrho = \frac{\sigma-1}{\sigma+1}$ denotes the contrast parameter, $f=\pi r_0^2$ the concentration of inclusions. 
They further developed the method of Rayleigh and applied it to various problems of the theory of composites \cite{Botten, Movchan, Nicorovici}.

\subsection{Method of Natanzon--Filshtinsky}
\label{se:ell}
The next important step in the mathematical treatment of the 2D composites was made by V.Ya. Natanzon \cite{Natanzon} in 1935 and  L.A. Filshtinsky (Fil'shtinskii) in 1964 (see papers \cite{Fil1,Fil2,Fil3}, his thesis \cite{Fil4}, the fundamental books \cite{Fil1970, Fil1991, Fil1994, Fil2001} and references in \cite{FilMit}). 
V.Ya. Natanzon and L.A. Filshtinsky modified and extended the method of Rayleigh to a 2D elastic doubly periodic problems but without reference to the seminal paper \cite{Rayleigh}. 

Rayleigh used the classic Laurent series \eqref{eq:Ray2a} in the domain $D$ of the unit cell and further periodically continued it by the Eisenstein summation \eqref{eq:s6}. The main idea of \cite{Natanzon} is based on the periodization of the complex potential $\varphi(z)$ without summation over the cells (we take the derivative as in \cite{Natanzon})
\begin{equation}
\varphi'(z)  = A_0+A_2 \wp(z)+ A_4 \wp''(z)+ A_6 \wp^{(iv)}(z)+\ldots.
 \label{eq:Ray2b1}
\end{equation} 
Filshtinsky used the series
\begin{equation}
\varphi(z)  = A_0 z -A_2 \zeta(z)- A_4 \zeta'(z)- A_6 \zeta'''(z)-\ldots,
 \label{eq:Ray2b}
\end{equation}
where $\zeta(z)$ and $\wp(z)=-\zeta'(z)$ are the Weierstrass elliptic functions. Application of series \eqref{eq:Ray2b1}-\eqref{eq:Ray2b} allows to avoid the study of the conditionally convergent lattice sum $S_2$. 

The Kolosov-Muskhelishvili formulae express the stresses and deformations in terms of two analytic functions. One of the functions has to be doubly periodic, hence, can be presented in the form \eqref{eq:Ray2b}. However, a combination of the first and second analytic functions satisfies periodic conditions which yields the following representation for the second function 
\begin{equation}
\Phi(z)  = B_0+B_2 \wp_1(z)+ B_4 \wp_1''(z)+\ldots,
 \label{eq:Ray3b}
\end{equation} 
where $\wp_1(z)$ is a new function introduced by Natanzon by means of the series
\begin{equation}
\wp_1(z) = -2 \sum_{(m_1,m_2) \in \mathbb Z^2 \backslash (0,0)} \left[ \frac{\overline{P}}{(z-P)^3} -  \frac{\overline{P}}{P^3} \right], 
 \label{eq:Ray4b}
\end{equation} 
$P=m_1 \omega_1+im_2 \omega_2$. Natanzon's function \eqref{eq:Ray4b} was systematically investigated in \cite{Fil1994, Natanzon}. In particular, simple expressions of $\wp_1(z)$ and its derivatives in terms of the Weierstrass elliptic functions were established. 

Substitution of the series \eqref{eq:Ray2b} into boundary conditions for the conductivity problem with one circular inclusion per cell yields the same Rayleigh's infinite system (see below \eqref{eq:ex9R4}) and low-order in concentration formulae for the effective conductivity. Filshtinsky \cite{Fil1,Fil3} obtained approximate analytical formulae for the local fields which can yield the effective conductivity after its substitution into \eqref{e11b}.

In  2012, Godin \cite{Godin} brought to a close the method of Natanzon-Filshtinsky for conductivity problems having repeated Filshtinsky's analytical approximate formulae for the local fields \cite{Fil1,Fil3}, used \eqref{e11b} and arrived at polynomials of order $12$ in $f$ for regular arrays. These polynomials are asymptotically equivalent to the approximation \eqref{proxr} established in 1978 and can be obtained by truncation from the series, first obtained in 1998, exactly written in the next section. The paper \cite{Godin} contains the reference to \cite{Fil1970} contrary to huge number of papers discussed in Sec.\ref{sec:exact}.

\subsection{Method of functional equations. Exact solutions}
\label{se:feq}
The methods of Rayleigh and of Natanzon--Filshtinsky are closely related to the method of functional equations \cite{Mit1993,MiR,GMN}. Roughly speaking, Rayleigh's infinite system is a discrete form of the functional equation. The method of functional equations stands out against other methods in exact solution to 2D problems with one circular inclusion per cell. Moreover, functional equations yield constructive analytical formulae for random composites \cite{GMN}.

In order to demonstrate the connection between the methods of Rayleigh and of functional equations we follow \cite{ryl} starting from the functional equation
\begin{equation}
\psi_1(z) = \varrho \sum_{m_1,m_2 \in \mathbb Z}\nolimits' 
\frac{r^2}{(z-P)^2} \;\overline{\psi_1 \left(\frac{r^2}{\overline{z-P}}\right)}+1,
\quad|z| \leq r_0.
 \label{eq:ex9R1}
\end{equation}
where $m_1,m_2$ run over integers in the sum $\sum'$ with the excluded term $m_1=m_2=0$. Here, $\psi_1(z)$ is the complex flux inside the disk $|z|<r_0$.
Let the function $\psi_1(z)$ be expanded in the Taylor series
\begin{equation}
\psi_1(z) = \sum_{m=0}^{\infty} \alpha_m z^m,
\quad|z| \leq r_0,
 \label{eq:ex9R2}
\end{equation}
Substituting this expansion into \eqref{eq:ex9R1} we obtain
\begin{equation}
\sum_{m=0}^{\infty} \alpha_m z^m = \varrho \sum_{m=0}^{\infty} r^{2(m+1)}\overline{\alpha_m}\sum_{m_1,m_2 \in \mathbb Z}\nolimits'
\frac{1}{(z-P)^{m+2}}+1,
 \label{eq:ex9R3}
\end{equation}
where the Eisenstein summation \eqref{eq:s6} is used. The function 
\begin{equation}
E_l(z)=\sum_{m_1,m_2 \in \mathbb Z}\frac{1}{(z-P)^{l}}
 \label{eq:exEi}
\end{equation} 
is called the Eienstein function of order $l$ \cite{weil}. Expanding every function $(z-P)^{-(m+2)}$ in the Taylor series and selecting the coefficients in the same powers of $z$ we arrive at the infinite $\mathbb R$-linear algebraic system
\begin{equation}
\alpha_l = \varrho \sum_{m=0}^{\infty} (-1)^m \frac{(l+m+1)!}{l!(m+1)!} S_{l+m+2}r^{2(m+1)}\overline{\alpha_m}+1, \;l=0,1,\ldots.
 \label{eq:ex9R4}
\end{equation}
This system \eqref{eq:ex9R4} coincides with the system obtained by Filshtinsky. Its real part is Rayleigh's system and Re $\alpha_1=\frac{\partial u_1} {\partial x_1} (0)$ (see \eqref{e11b}).

The functional equation \eqref{eq:ex9R1} is a continuous object more convenient for symbolic computations than the discrete infinite systems \eqref{eq:ex9R4}. 
Instead of expansion in the discrete form described above we solve the functional equation by the method of successive approximations uniformly convergent for any $|\varrho|\leq 1$ and arbitrary $r_0$ up to touching \cite{GMN}. Application of \eqref{e11b} yields the exact formula for the effective conductivity tensor
\begin{align}
&\sigma_{11}- i \sigma_{12} =1+2\varrho f+2\varrho^2 f^2 \frac{S_2}{\pi} + \label{eq:ex10} 
\\
 &\quad\frac{2\varrho^2 f^2}{\pi} \sum_{k=1}^{\infty}\varrho^{k}
\sum_{m_1=1}^{\infty} \sum_{m_2=1}^{\infty}\cdots
\sum_{m_k=1}^{\infty} s_{m_1}^{(1)} s_{m_2}^{(m_1)}
\ldots s_{m_k}^{(m_{k-1})}s_{1}^{(m_{k})}
\left(\frac{f}{\pi} \right)^{M-k}, 
\notag
\end{align} 
where $M=2(m_1+m_2+\dots+m_k)$ and
\begin{equation}
s_k^{(m)} = \frac{(2m+2k-3)!}{(2m-1)!(2k-2)!} S_{2(m+k-1)}.
 \label{eq:2.11}
\end{equation}
The Eisenstein--Rayleigh lattice sums $S_m$ are defined as $S_m=\sum_{P\neq 0} P^{-m}$
and can be determined by computationally effective formulae \cite{GMN}. The component $\sigma_{22}$ is calculated by \eqref{eq:ex10} where $S_2$ is replaced by $2\pi - S_2$. 

Formula is exact and first was described in 1997 in the papers \cite{Mit1997} in the form of expansion on the contrast parameter $\varrho$. Justification of the uniform convergence for any $|\varrho|\leq 1$ and for an arbitrary radius up to touching disks was established in \cite{Mit1997a,Mit1997}.  The papers \cite{Mit1998, Mit2007} contains transformation of the contrast expansion series from \cite{Mit1997} to the series \eqref{eq:ex10} more convenient in computations.

The relation  $S_2=\pi$ holds for the square and hexagonal arrays (macroscopically isotropic composites). It is worth noting that in this case the terms with $m_1=m_2= \ldots = 1$ in the sum over $k$ and the first three terms in \eqref{eq:ex10} form a geometric series transforming into the Clausius-Mossotti approximations (Maxwell's formula)
\begin{equation}
\label{eq:CMA}
\sigma_e \approx 
\frac{1+ f \varrho}{1- f \varrho}.
\end{equation} 

The series \eqref{eq:ex10} can be investigated analytically and numerically. For instance, for the hexagonal array of the perfectly conducting inclusions ($\varrho=~1$) it can be written in the form
\begin{eqnarray}
\label{series}
\sigma_e{(f)}=&&
1 + 2 f + 2 f^2 + 2 f^3 + 2 f^4 +2 f^5 +2 f^6
\nonumber\\
&+&
2.1508443464271876 f^7 +2.301688692854377 f^8
\nonumber\\
 &+& 
2.452533039281566 f^9 +2.6033773857087543 f^{10}
 \nonumber\\
 &+&
2.754221732135944 f^{11} +2.9050660785631326 f^{12} 
 \nonumber\\
 &+&
3.0674404324522926 f^{13}  +3.2411917947659736 f^{14} 
 \nonumber\\
&+&
3.426320165504177 f^{15} +3.6228255446669055 f^{16}
\nonumber\\
&+& 
3.8307079322541555 f^{17} +4.049967328265928 f^{18} 
\nonumber\\ 
 &+&
 4.441422739726373 f^{19} +4.845994396051242 f^{20} 
 \nonumber\\ 
 &+&
5.264540375940583 f^{21} +5.69791875809444 f^{22} 
 \nonumber\\ 
&+& 
6.146987621212864 f^{23} +6.6126050439959 f^{24}
\nonumber\\ 
 &+&
7.135044602470776 f^{25}+7.700073986554016 f^{26} 
 \nonumber\\ 
&+& 
O(f^{27}).
\end{eqnarray}
The first 12 coefficients in the expansion of \eqref{proxr} in the series in $f$ for $\varrho=1$ coincide with the coefficients of \eqref{series}.   

Application of asymptotic methods to \eqref{series} yields analytical expressions near the percolation threshold  \cite{GMN}
\begin{equation}
\label{fin4}
\sigma_e(f)=\alpha(f) \;\frac{F(f)}{G(f)},
\end{equation} 
where 
\begin{equation}
\label{fin}
\begin{array}{llll}
\alpha(f)=
\frac{4.82231}{\left(\frac{\pi}{\sqrt{12}}-f\right)^{\frac 12}}-5.79784+
\\
2.13365 \left(\frac{\pi}{\sqrt{12}}-f\right)^{\frac 12}-0.328432 \left(\frac{\pi}{\sqrt{12}}-f\right),
\end{array}
\end{equation}
\begin{equation}
\label{fin24}
\begin{array}{llll}
F(f)=
1.49313+1.30576 f+0.383574 f^2+0.467713 f^3+
\\
0.471121 f^4 + 0.510435 f^5 + 0.256682 f^6 +
0.434917 f^7+\\
0.813868 f^8+0.961464 f^9+
0.317194 f^{10}+0.377055 f^{11}-\\1.2022 f^{12}-0.931575 f^{13}
\end{array}
\end{equation} 
and
\begin{equation}
\label{fin34}
\begin{array}{llll}
G(f)=
1.49313+1.30576 f+0.383574 f^2+0.394949 f^3+
\\
0.4479 f^4+0.5034 f^5+0.3033 f^6+
0.2715 f^7+0.7328 f^8+\\0.827239 f^9+
0.25509 f^{10}+0.239752 f^{11}-1.26489 f^{12}-f^{13}.
\end{array}
\end{equation}
The function \eqref{fin4} is asymptotically equivalent to the polynomial \eqref{series}, as $f\to 0$, and to Keller's type formula \cite{BN}
\begin{equation}
\label{fin40}
\sigma_e(f)=\frac{\sqrt[4]{3} \pi^{\frac 32}}{\sqrt{2}}  \frac{1}{\sqrt{\frac{\pi}{\sqrt{12}}-f}}, \quad \mbox{as} \; f \to f_c=\frac{\pi}{\sqrt{12}}.
\end{equation} 

\begin{figure}[ht]
\centering
\includegraphics[width=0.8\textwidth]{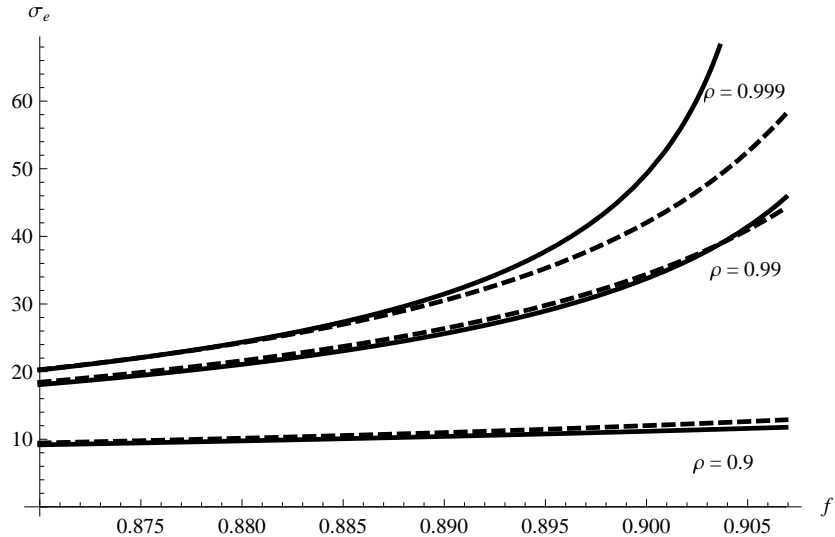}
\caption{Effective conductivity for the hexagonal array $\sigma_e(f,\varrho)$ calculated with \eqref{symmet11}-\eqref{symmethexW} (solid line) and \eqref{proxr} (dashed line).}
\label{Fig-McPh}
\end{figure}

The series \eqref{eq:ex10} for the hexagonal array when $|\varrho|\leq 1$ yields 
\begin{align}
\label{Phex}
\sigma_e(f,\varrho) = \frac{1+\varrho f}{1-\varrho f}+
0.150844\varrho^3 f^7+0.301688\varrho^4 f^8 +0.452532\varrho^5 f^9+
\\
\notag
0.603376\varrho^6 f^{10}+0.75422\varrho^7 f^{11}+\cdots .
\end{align}
The asymptotic analysis of the local fields when  $|\varrho|\to 1$ and $f\to f_c$ was performed in \cite{ryl2} where a criterion of the percolation regime was given, i.e., the domain of validity of  \eqref{Phex}. The following formula was established in \cite{GMN}
\begin{equation}
\label{symmet11}
\sigma_e({f,\varrho})=\sigma^{*}({f,\varrho}) \; 
\frac{U(f,\varrho)}{W(f,\varrho)},
\end{equation}
where
\begin{equation}
\label{symmethex}
\sigma^{*}({f,\varrho})=\left( 1+\frac{\varrho f}{f_c}\right)^{\frac 12} \left( 1-\frac{\varrho f}{f_c}\right)^{-\frac 12},
\end{equation}
\begin{equation}
\begin{array}{ccccc}
U(f,\varrho)=-\varrho^{11} f^{11}-0.660339\varrho^9 f^9+\varrho^7 (1.44959 f^4-0.232667) f^7+\\
\varrho^5 (1.59231 f^9+0.457218 f^5)+\varrho^3 (1.99365 f^7+2.10888 f^3)+
\\
11.8598\varrho f+26.4332
\end{array}
\end{equation}
and
\begin{equation}
\label{symmethexW}
\begin{array}{ccccc}
W(f,\varrho)=
\varrho^7 f^7 (0.232667-1.44959 f^4)+ \varrho^{11} f^{11}+0.660339 \varrho^9 f^9+\\
\varrho^5 (-1.59231 f^9-0.457218 f^5)+\varrho^3(-1.99365 f^7-2.10888 f^3)-
\\
11.8598 \varrho f+26.4332.
\end{array}
\end{equation} 
The function \eqref{symmet11} is asymptotically equivalent to \eqref{Phex}, as $(\varrho f) \to 0$, and to the percolation regime. Formulae \eqref{symmet11}-\eqref{symmethexW} and \eqref{proxr} for high concentrations are compared in Fig.\ref{Fig-McPh}.

\section{"Exact solutions"}
\label{sec:exact}
Perhaps, the first too magnified usage of the term "exact solution" in the theory of composites began in 1971 with  Sendeckyj's paper \cite{Sendeckyj} in the first issue of Journal of Elasticity where "an exact analytic solution is given for a case of antiplane deformation of an elastic solid containing an arbitrary number of circular cylindrical inclusions". Actually, as it is written in this paper, a method of successive approximations based on the method of images was applied to find an approximate solution in analytical form. Sendeckyj's method is a direct application of the generalized alternating method of Schwarz developed by Mikhlin \cite[Russian edition in 1949]{Mikhlin} to circular finitely connected domains. This method does not always converge. The method of Schwarz was modified in \cite{Mit1993} where exact\footnote{not "exact"} solution of the problem was obtained in the form of the Poincar\'{e} type series \cite{Mit2014}. Its relation to the series method leading to infinite systems is described in \cite{Mit1995}.\newline

In the book \cite[Introduction]{Kushch}, Kushch asserts that his solutions are "complete" and "the exact, finite form expressions for the effective properties" obtained. The same declarations are given in \cite{Kushch2008, Kushch2010}. For instance, "exact" and "complete solution of the many-inclusion problem" is declared in \cite{Kushch2006}. Below \cite[Sec.5]{Kushch2006} the authors explain that "solution we have derived is asymptotically exact. It means that to get the exact values, one has to solve a whole infinite set of linear equations". In \cite[Sec.9.2.2]{Kushch}, difficulties in solution to infinite systems arisen for a finite number of disks are described. 
In the paper \cite{Kushch2010} devoted to the same problem, Kushch writes that "an exact and finite form expression of the effective conductivity tensor has been found". However, the "exact" formula (62) from \cite{Kushch2010} contains parameters $F_n^{(p)}$ which should be determined numerically from an infinite system. 

As it is explained in Introduction to the present paper, a regular (in the sense of \cite{KK}) infinite system of linear algebraic equations is a discrete form of the continuous Fredholm's integral equation. The truncation method is applied to the both discrete and continuous Fredholm's equations refers to numerical methods in applied mathematics, since a special data set of geometrical and physical parameters is usually taken to get a result. Then, following logical reasoning and Kushch's declarations we should use the term "exact solution" for numerical solutions  obtained from Fredholm's equations that essentially distorts the sense of exact solution. 

\begin{figure}[ht]
\centering
\includegraphics[width=0.8\textwidth]{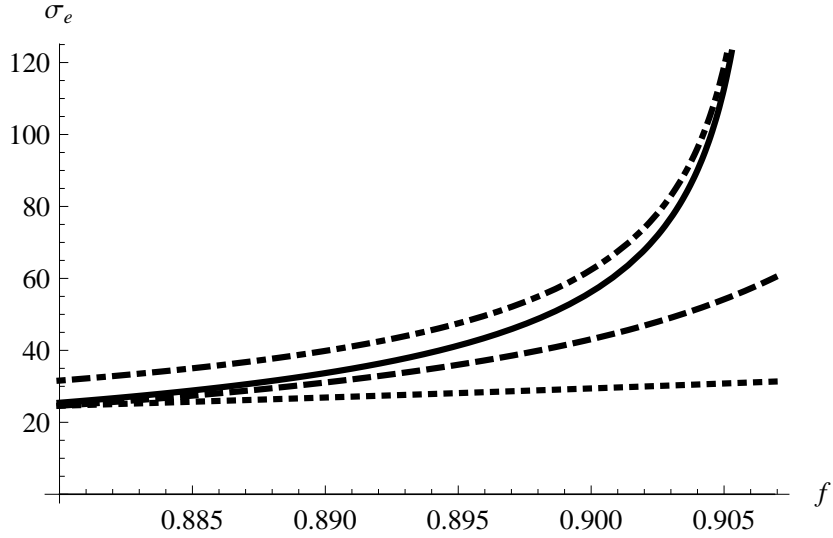}
\caption{Effective conductivity for the hexagonal array. Different expressions for $\sigma_e$ calculated with \eqref{fin4}-\eqref{fin34} (solid line), \eqref{proxr} (dashed line), \eqref{fin40} (dot-dashed line). Formula \cite[(21) and (27)]{Balagurov3} is presented by dotted line. 
}
\label{Fig-Balagurov}
\end{figure}

Another question concerns the declaration "complete solution" \cite{Kushch,Kushch2006,Kushch2008,Kushch2010}. The completely solved problem should not be investigated anymore. Such a declaration by Kushch ignores exact solutions described in Sec.\ref{se:feq} including the exact formulae obtained before, c.f. \cite{Mit1993}. It is demonstrated in Fig.\ref{Fig-Balagurov} that the "complete solution" by Kushch is no longer complete for high concentrations. \newline  

Beginning from 2000 Balagurov \cite{Balagurov1}-\cite{Balagurov5} has been applying the method of Natanzon and Filshtinsky described above in Sec.\ref{se:ell} (after exact solution to the problem for regular array of disks in 1997-1998, see Sec.\ref{se:feq}) without references to them. The main difference between \cite{Balagurov1}-\cite{Balagurov5} and \cite{Natanzon}, \cite{Fil1970, Fil1991, Fil1994, Fil2001} lies in the terminology. Balagurov used the terminology  of conductivity governed by Laplace's equation (harmonic functions).  Natanzon and Filshtinsky used the terminology of elasticity and heat conduction governed by bi-Laplace's (biharmonic functions) and by Laplace's equation, respectively. Moreover, Filshtinsky separated in his works plane and antiplane elasticity which is equivalent to a separate consideration of bi-Laplace's and Laplace's equation.

For instance, the paper \cite{Balagurov3} is devoted to application of the Natanzon-Filshtinsky representation \eqref{eq:Ray2b} to the square array of circular cylinders. Exactly having repeated \cite{Natanzon,Fil1970, Fil1991, Fil1994, Fil2001}, Balagurov and Kashin obtained the complex infinite system \eqref{eq:ex9R4} written in the paper \cite{Balagurov3} as (A6). Further, Balagurov and Kashin \cite[formulae (21) and (27)]{Balagurov3} write the effective conductivity up to $O(f^{13})$. This formula from \cite{Balagurov3} is asymptotically equivalent up to $O(f^{13})$  to  formula (14) obtained in the earlier paper \cite{Mc}. Balagurov's results are not acceptable for high concentrations as displayed in Fig.\ref{Fig-Balagurov}. \newline 

Parnell and Abrahams \cite{Abrahams} derived "new expressions for the effective
elastic constants of the material ... in simple closed forms (3.24)-(3.27)".
They "not appealed directly to the theory of Weierstrassian
elliptic functions in order to find the effective properties". Actually, they repeated fragments of Eisenstein's approaches (1848) in \cite[Sec.4.1]{Abrahams}. 
Therefore, the same Natanzon-Filshtinsky method was applied to regular arrays of circular cylinders in \cite{Abrahams}.\footnote{The most unexpected paper concerning doubly periodic functions belongs to Wang \& Wang \cite{Yufeng} who in this one paper i) rediscovered Weierstrass's functions \cite[(2.29) and (2.30)]{Yufeng}, ii) rediscovered Eisenstein's  summation approach \cite[(2.18)]{Yufeng}, and iii) solved a boundary value problem easily solved by means of the standard conformal mapping.}\newline

A series of papers beginning form 2000, see \cite{Sabina1,Sabina2,Sabina3} and references therein, contains a wide set of "exact formulae" for the effective constants hardly accepted as constructive following the lines of Sec.\ref{sec:Int}. Some of them \cite{Sabina1, Guinovart, Sabina1c} have the misleading title "Closed-form expressions for the effective coefficients ...". In the paper \cite{Sabina3}, "the local problems are solved for the case of fiber reinforced composite and the exact-closed formulae for all overall thermoelastic properties" etc. The main results are obtained by the method of Natanzon--Filshtinsky (1935, 1964) without references to it. As in the previous papers, the series \eqref{eq:Ray2b1}-\eqref{eq:Ray2b} are substituted into the boundary conditions and Rayleigh's type system of linear algebraic equations is obtained. 

Let us consider the recent paper \cite{Nascimento} where the 2D square array of
circular cylinders and the 3D simple cubic array of spheres of two different radii $R_s$ and $R_l$ were considered when the ratio $R_s$ to $R_l$ was infinitesimally small. The authors declare that "the interaction of periodic multiscale heterogeneity arrangements is exactly accounted for by the reiterated homogenization method. The method relies on an asymptotic expansion solution of the first principles applied to all scales, leading to general rigorous expressions for the effective coefficients of periodic heterogeneous media with multiple spatial scales". This declaration is not true, because a large particle does not interact with another particle of vanishing size. An effective medium approximation valid for dilute composites \cite{Mit2013} was actually applied in \cite{Nascimento} as the reiterated homogenization theory. As a consequence, formulae (18) and (19) from \cite{Nascimento} can be valid to the second order of concentration \cite{Mit2013} and additional numerically calculated "terms" are out of the considered precision. The considered problem refers to the general polydispersity problem discussed in 2D statement in \cite{BiM2005}. It is not surprisingly that the description of the polydispersity effects in \cite{BiM2005} and \cite{Nascimento} are different since the "exact" formula from \cite{Nascimento} holds only in the dilute regime. This is the reason why the effective conductivity is less than the conductivity of matrix reinforced by higher conducting inclusions in Figs 3 and 7 of \cite{Nascimento} for high concentrations.

\section{Conclusion}
\label{sec:disc}
We can draw the following conclusions. 

1. Authors' claims on closed-form expressions for the effective coefficients, true for any parameter values, are absolutely unjustified.

2. The results obtained by them in this particular case are correct, but they have a very limited field of applicability and are presented in a form that does not allow direct use. Only after making considerable efforts and becoming, in fact, the co-author of the paper, the reader is able to get simple formulae. As a result, reader is convinced that this is a minor modification (and not in the direction of improvement) of well-known formulae.

The question arises of the usefulness of such works - they are clearly inaccessible to the engineer, they do not contain new mathematical results, the field of applicability has not been evaluated in any way.

The general conclusion is that the expressions
"exact" or "closed-form" solutions are very obligatory. The authors of the articles should not use them in vain, and reviewers and editors must stop unreasonable claims.

Note also numerous attempts to reinvent the wheel, which look especially strange in time of Google's Empire and scientific social networks (Research Gate, etc.)

In fact, an infinite system of linear algebraic equations contains a remarkable amount of information, which the investigator should do his best to extract (see, c.f., \cite{KK}). However, we can not usually see this in papers with "exact solutions".

We note one more aspect of the application of analytic, in particular, asymptotic, methods. The significant disadvantage inherent in them - the local nature of the results obtained - is overcome by modern methods of summation and interpolation (Pad\'{e} approximants, two--point Pad\'{e} approximants, asymptotically equivalent functions, etc. \cite{Andrianov2a,GMN}). Example from our paper: formulae (2.24)-(2.27) are deduced by asymptotic matching (2.23) near $f = 0$ and (2.28) near $f = f_c$.
%

\section*{Acknowledgments}
Authors thanks Dr Galina Starushenko for fruitful discussions. 

\end{document}